\documentclass[cits]{PoS} 
\usepackage{pstricks,pst-plot,pst-node,pst-coil}
\usepackage{graphicx}
\usepackage{amsmath}
\usepackage{amssymb}
\usepackage{wrapfig}
\usepackage{enumerate}

\newcommand{\dirFig}{.}  

\newcommand{\mytitle}{
Lattice study on two-color QCD with six flavors of dynamical quarks
}

\title{
\mytitle
} 

\ShortTitle{
\mytitle
} 

\author{\speaker{M.~Hayakawa}$^{a}$, K.-I.~Ishikawa$^{bc}$,
Y.~Osaki$^{b}$, S.~Takeda$^{d}$
and N.~Yamada$^{e,f}$\\
\\
\llap{$^a$}Department of Physics, Nagoya University, 
Nagoya 464-8602, Japan\\
\llap{$^b$}Department of Physics, Hiroshima University, 
Higashi-Hiroshima 739-8526, Japan\\
\llap{$^c$}RIKEN Advanced Institute for Computational Science,
Kobe, Hyogo 650-0047, Japan\\
\llap{$^d$}School of Mathematics and Physics, 
College of Science and Engineering, Kanazawa University, \\
Kakuma-machi, Kanazawa, Ishikawa 920-1192, Japan\\
\llap{$^e$}High Energy Accelerator Research Organization {\rm (}KEK{\rm )}, 
Tsukuba, Ibaraki 305-0801, Japan\\
\llap{$^f$}The Graduate University for Advanced Studies, 
Tsukuba, Ibaraki 305-0801, Japan\\ 
} 

\abstract{
  We study the dynamics of
${\rm SU(2)}$ gauge theory with $N_F=6$ Dirac fermions 
by means of lattice simulation to investigate if 
they are appropriate to 
realization of electroweak symmetry breaking.
 The discrete analogue of beta function
for the running coupling constant
defined under the Schr\"{o}dinger functional boundary condition 
are computed on the lattices up to linear size of $L/a=24$ 
and preclude the existence of infrared fixed point below $g^2 \sim 7.6$.
 Gluonic observables such as heavy quark potential, string tension,
Polyakov loop suggest that the target system is in the confining phase
even in the massless quark limit.
}

\FullConference{The 30th International Symposium on Lattice Field Theory\\
		 July 24-29, 2012\\
		 Cairns, Australia}

\begin{document}


 After the seminal work \cite{Appelquist:2007hu},
great attention has been paid to the possibility to study
nearly conformal dynamics of gauge theory by means of lattice
simulation,
which are expected to trigger electroweak symmetry breaking.
 So far, many works have been involved in the calculations
on ${\rm SU(3)_C}$ gauge theory with $N_F$ Dirac fermions
in the fundamental representation. 
 Recently, it was pointed out that three-color QCD with ten-flavors
may be conformal in the infrared limit with large mass anomalous dimension
\cite{Appelquist:2012nz,Hayakawa:2012:prep}.
 However, large volume simulations indicated that
chiral symmetry breaking occurs in the twelve-flavor system
\cite{Fodor:2009wk}.
 The large scale simulation via 
Monte-Carlo renormalization group \cite{Hasenfratz:2011xn}
will be effective to extract the genuine dynamics of the system.

 Here we focus on a series of ${\rm SU(2)_C}$ gauge theories
with $N_F$ Dirac fermions in the fundamental representation,
and try to find out a gauge system with the critical number of flavors
$N_F^{\rm crt}$,
at which the chiral condensate $\overline{\psi} \psi$ gets
large anomalous dimension.
 ${\rm SU(2)}$ has a symplectic form 
so that its fundamental representation is pseudo-real,
and that the chiral symmetry of $N_F$-flavor system 
is enhanced to ${\rm SU}(2N_F) \supset
{\rm SU}(N_F)_{\rm L} \times {\rm SU}(N_F)_{\rm R} \times {\rm U(1)_V}$.
 If chiral symmetry is spontaneously broken,
the plausible unbroken symmetry subgroup is ${\rm SP}(2 N_F)$. 
 But, this enhanced unbroken symmetry
may contain the electroweak symmetry
$G_{\rm W} = {\rm SU}(2)_{\rm L} \times {\rm U(1)_Y}$
depending on the representation of new ``quarks'' under $G_{\rm W}$.
 It is thus inevitable to examine the vacuum alignment issue 
\cite{VacuumAlignment}, 
\emph{i.e.}, whether $G_{\rm W}$ is broken or not.
 The transition between confinement and deconfinement is argued to be
second-order in pure Yang-Mills theory with ${\rm SU(2)_C}$
\cite{Svetitsky:1982gs}
while it is first-order for ${\rm SU(3)_C}$.
 Therefore, ${\rm SU(2)_C}$ chiral dynamics
can differ even qualitatively
from those of ${\rm SU(3)_C}$, in particular at $N_F^{\rm crt}$.

 Actually, two-color QCD has also been studied thus far.
 Iwasaki \emph{et al.} showed that $N_F = 3$ system is conformal
in the infrared (IR) limit \cite{Iwasaki:2003de}
through study of phase structure of Wilson fermions,
while the perturbatively calculated $\beta$ function
suggests that $6 \le N_F^{\rm crt} \le 8$.
 Afterwards, running gauge coupling constant has been 
calculated nonperturbatively for
two-color QCD
with six-flavors \cite{Bursa:2010xn,Voronov:2011zz,Karavirta:2011zg},
and eight-flavors \cite{Ohki:2010sr}, 
implying that those systems are conformal in the IR limit
\cite{Bursa:2010xn,Karavirta:2011zg,Ohki:2010sr}.

 We note that 
${\rm SU(2)_L}$ gauge theory with three generations of quarks and leptons
is exactly the system of our concern here, 
the two-color QCD with six massless Dirac fermions.
 Our question is if
the quantum-mechanical dynamics of the presumed copy of 
the existing gauge symmetry and fermionic matters 
but with large $\Lambda_{\rm ``QCD''}$ 
could play the role of 
triggering spontaneous breakdown
of the electroweak symmetry, ${\rm SU(2)_L \times U(1)_Y}$. 

 The purpose of this article is to report our first result for
the dynamical features of two-color QCD with six-flavors 
according to the simulation in the framework of lattice gauge theory.
 Throughout this work,
the standard Wilson plaquette gauge action
with unimproved Wilson fermions is used for simulation.

 We measure $g_{\rm SF}(L/a,\,g_0^2)$ 
for sets of $(L/a,\,g_0^2 = \frac{4}{\beta})$ defined under
the Schr\"{o}dinger functional boundary condition
with the twist angle for the quark fields set to $0$
\cite{SFcoupling}.
 Compared to the preceding work \cite{Bursa:2010xn},
computation on larger lattices, 
$L/a = 6$, $8$, $12$, $16$, $18$ and $24$, 
is done with fine tuning of the critical value of hopping parameter
for every pair of $(L/a,\,g_0^2)$.
 Data are fit to the presumed functional form
\begin{align}
 \frac{g_0^2}{g_{\rm fit}^2(L/a,\,g_0^2)} 
 &=
 \frac{1 - a_{L/a}^{(1)}\,g_0^4}
  {\displaystyle{
   1 + p_{L/a,\,1} g_0^2
   + \sum_{k=2}^N
   a_{L/a}^{(k)}\,g_0^{2k}
  }} \,,
\end{align}
where the coefficient $p_{L/a,\,1}$ is obtained by the one-loop calculation.
 Using the fit result $g_{\rm fit}^2(l,\,g_0^2)$, the discrete beta function
\cite{Shamir:2008pb}
\begin{align}
 B^{(s)}(u;\,
    l_1 \mapsto l_2 = s l_1)
 &\equiv
 \left.
  \frac{1}{g^2_{\rm fit}(l_2 = s l_1,\,g_0^2)}
 \right|_{u = g_{\rm fit}^2(l_1,\,g_0^2)}
 - 
 \frac{1}{u}\nonumber\,,
\end{align}
for a fixed step scaling factor $s$ can be calculated.

\begin{wrapfigure}[15]{r}{7.4cm}
\raisebox{-11.5em}[0pt][11.5em]{
\includegraphics[width=7.4cm,clip]{\dirFig/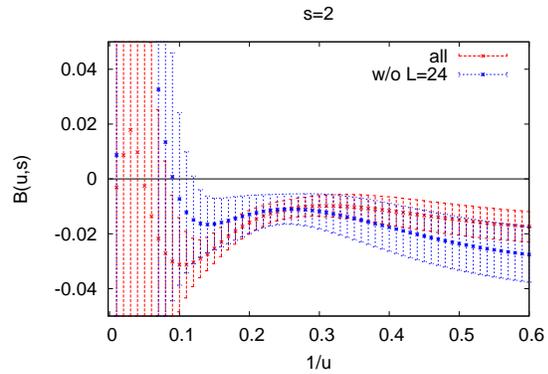}
}
\caption{
Discrete beta function for the step scaling $s=2$.
 The horizontal axis is
the {\it inverse} of squared renormalized coupling constant.
 Red plots were obtained from all data,
while blue plots were obtained from data without $L/a=24$.}
\label{fig:SFcouplig}
\end{wrapfigure}

 Figure \ref{fig:SFcouplig}
shows our result for its continuum limit $B\left(u,\,s = 2\right)$ 
taken in the same manner as in Ref.~\cite{Hayakawa:2010yn}.
 The result precludes the existence of infrared fixed point $u_\star$
below $7.6$.
 We are also trying to
calculate the anomalous dimension of the chiral condensate 
$\overline{\psi} \psi$. 
 We find that the result is roughly consistent with
the perturbative prediction and the 
systematic uncertainties are under examination.

\begin{figure}[htb]
 \begin{minipage}[t]{.48\textwidth}
  \includegraphics[width=7.3cm,clip]{\dirFig/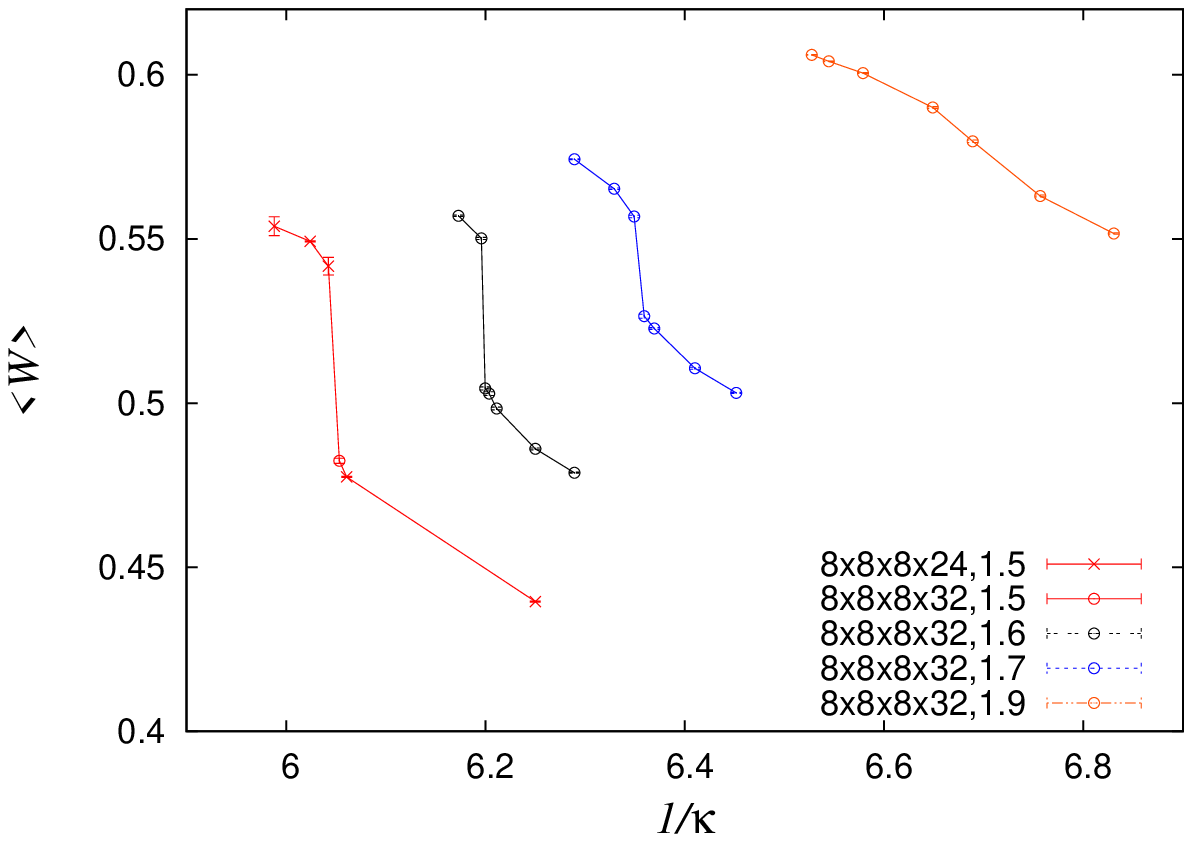}
  \caption{
  Plaquette, $\left<W\right>$, 
  versus $1 /\kappa$ on the lattices of spatial linear size 
  $L/a=8$ with $\beta=1.5$, $1.6$, $1.7$ and $1.9$.}
  \label{fig:plaquette_8x8x8}
 \end{minipage}
 \hfill
 \begin{minipage}[t]{.48\textwidth}
  \includegraphics[width=7.3cm,clip]{\dirFig/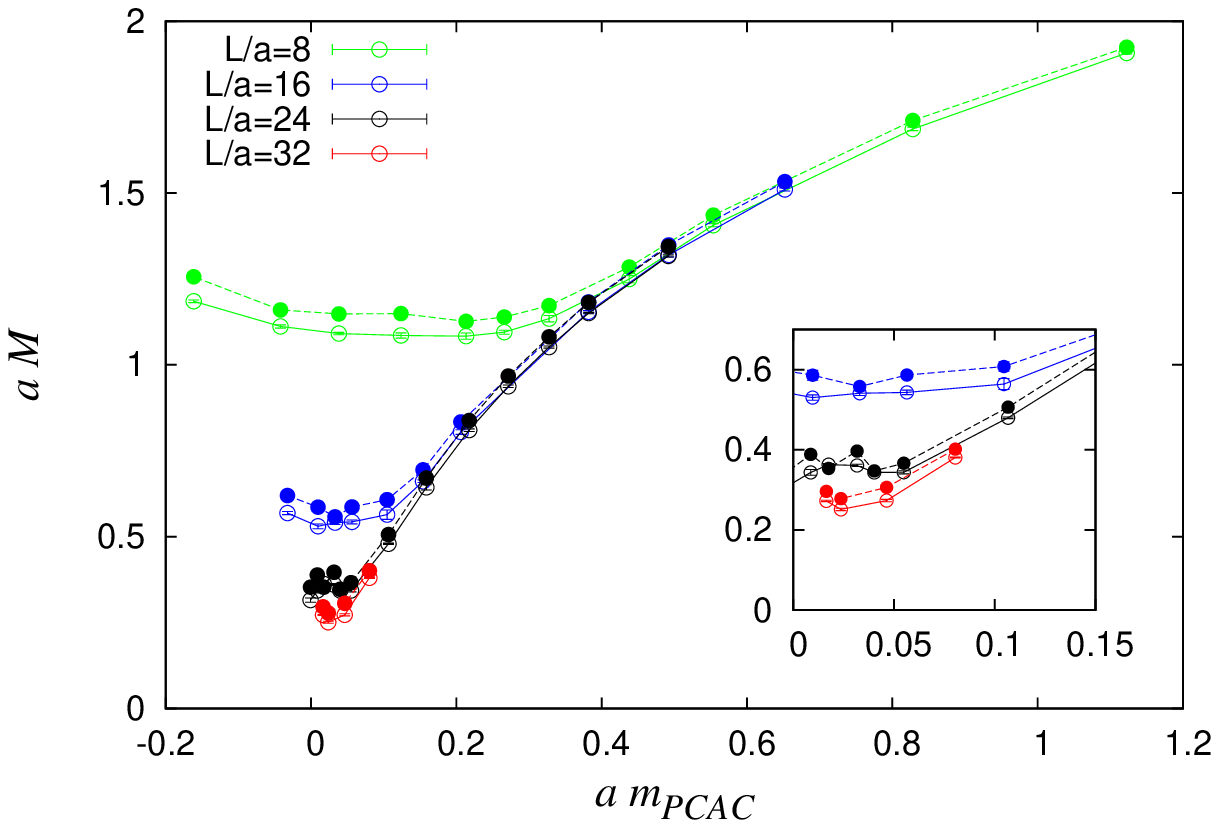}
  \caption{Lightest pseudoscalar ($\circ$) and vector ($\bullet$)
   meson masses versus
   PCAC mass $m_{\rm PCAC}$ in 
   lattice unit at $\beta=2.0$ on four different volumes.}
  \label{fig:ps_v_versus_pcac}
 \end{minipage}
\end{figure}

 The above result suggests that $N_F=6$ is not far from $N_F^{\rm crt}$
and motivates us to investigate the theory
further from a different point of view.
 We thus study the dependence of mesonic spectrum on quark masses.
 First, in order to see the phase structure as a statistical system 
and fix the simulation parameters, we performed a 
scan on the $\left(\beta,\,\kappa\right)$-plane with 
relatively small lattices.
 Figure \ref{fig:plaquette_8x8x8} shows 
the plaquette $\left<W\right>$ as a function of $1/\kappa$ 
on $8^3 \times 24$ and $8^3 \times 32$.
 For $\beta \lesssim 1.7$,
there is a region of $\kappa$ in which $\left<W\right>$ changes rapidly.

 Thus, we choose $\beta = 2.0$ to simulate larger lattices 
for our target system.

\begin{figure}[htb]
 \begin{minipage}[t]{.48\textwidth}
  \includegraphics[width=7.3cm,clip]{\dirFig/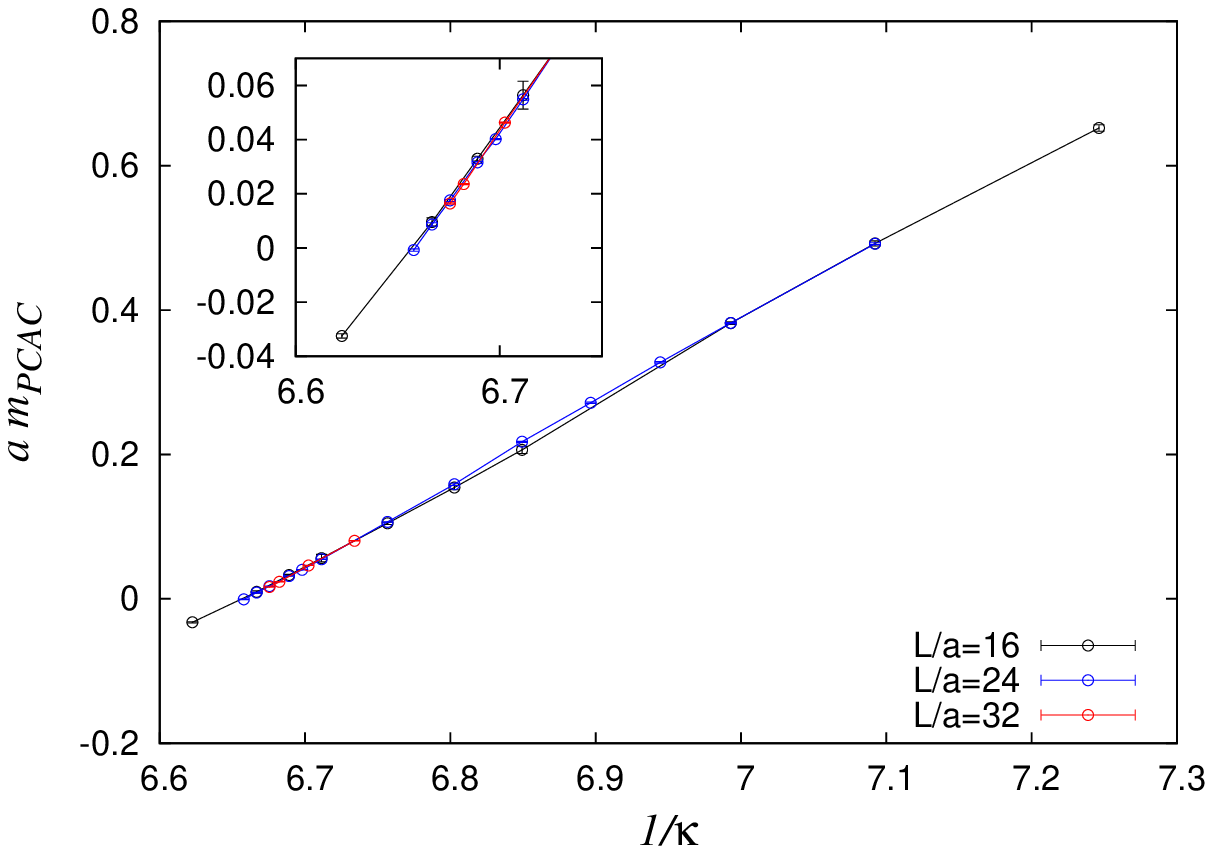}
  \caption{PCAC mass $m_{\rm PCAC}$ versus 
   $1/\kappa$ at $\beta=2.0$. 
  The inset zooms in the vanishing $m_{\rm PCAC}$ region.
  Finite size effect is negligibly small.}
  \label{fig:kappaInv_versus_pcac}
 \end{minipage}
 \hfill
 \begin{minipage}[t]{.48\textwidth}
  \includegraphics[width=7.5cm,clip]{\dirFig/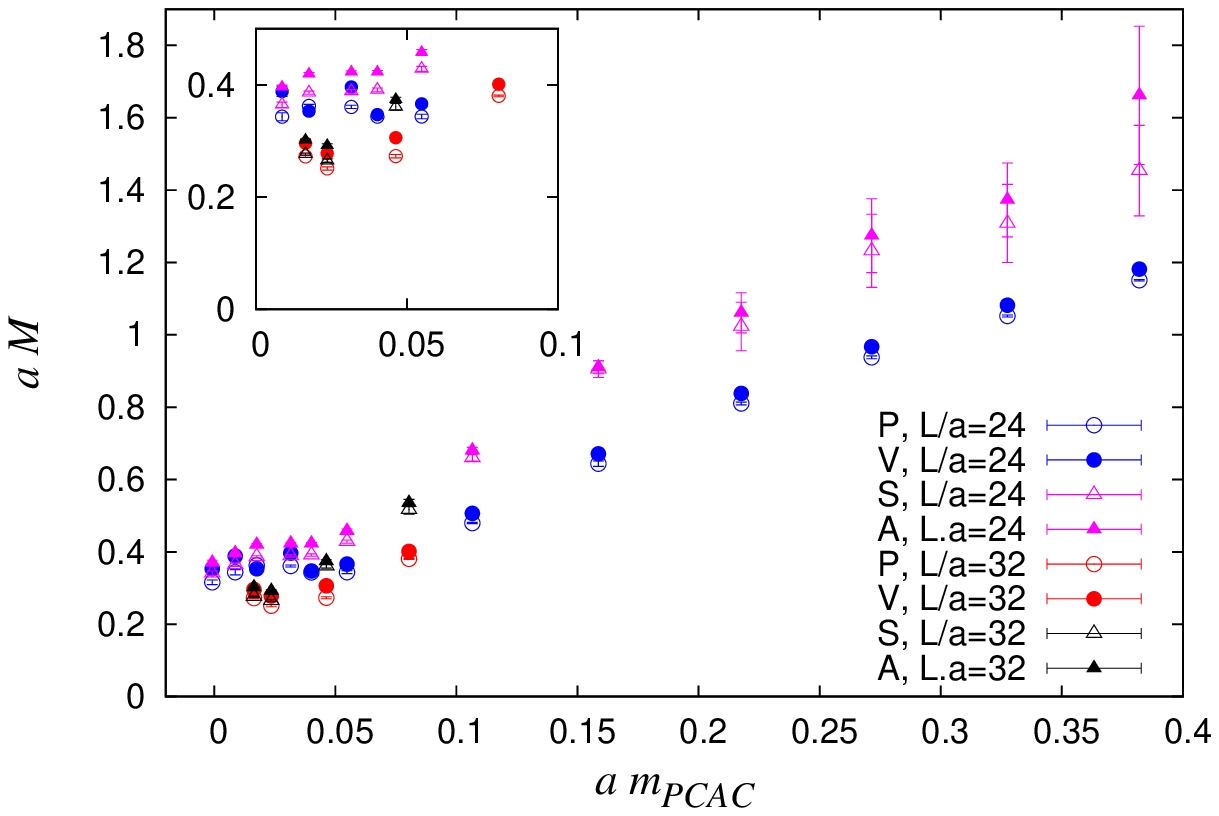}
  \caption{Masses of lightest pseudoscalar (P), vector (V), scalar (S) and
   axial-vector (A) mesons for $L/a=24$ and $L/a=32$.}
  \label{fig:allMesonsVesusPcacMass}
 \end{minipage}
\end{figure}

 Figure \ref{fig:ps_v_versus_pcac} shows
the masses of the lightest pseudoscalar and vector mesons, 
where the dependence on $\kappa$ is translated to
that on $a m_{\rm PCAC}$ 
via Figure \ref{fig:kappaInv_versus_pcac} 
and the following properties are observed:
\begin{enumerate}[{(}1{)}]
\item
 The meson masses are bounded from below at small $a m_{\rm PCAC}$.
 The lower bound $a m_\psi^{\rm sat}(L/a)$ depends on $L/a$.
\label{item:MesonMassBounded}
\item
  Even for small $a m_{\rm PCAC}$,
the ratio of the mass of the vector meson to that of the pseudoscalar meson 
is not far from $1$.
 Figure \ref{fig:allMesonsVesusPcacMass} shows
that the scalar meson is paired with the axial-vector meson
\footnote{
 The fit is performed
for the two-point correlation functions 
in the scalar and axial-vector channels
before substantial fluctuation sets in 
at large $t/a$. 
}, 
and that their masses are well above those of
the lightest pseudoscalar and vector mesons 
before saturation occurs.
\label{item:Degeneracy_P_V}
\end{enumerate}
 For $L/a=8$,
we also checked that the property $(\ref{item:MesonMassBounded})$ persists
for Iwasaki gauge action and/or clover fermions, and that
$a m_\psi^{\rm sat}(L/a=8)$ depends modestly on the types of action and $\beta$.
 This saturation phenomenon was also observed 
in ${\rm SU(3)_C}$ gauge theory
with two sextet quarks \cite{DeGrand:2008kx}.
 When we look at Figure \ref{fig:ps_v_versus_pcac} 
from large $a m_{\rm PCAC}$,
the meson masses branch at larger $a m_{\rm PCAC}$ 
with $a m_\psi^{\rm sat}(L/a)$ for smaller $L/a$, 
which strongly indicates
that this saturation originates from finite size effect. 
 
 We comment on
the dependence of $a m_\psi^{\rm sat}(L/a)$
on the linear size $L/a$ of the system.
 The spatial correlation length $\xi_\psi$ 
cannot become larger than the system size $L/a$ 
and saturated at $\sim L/a$.
 Thus, 
{\it if} the Compton wavelength 
$2 \pi / (a m_\psi)$ coincides with $\xi_\psi$, 
$a m_\psi^{\rm sat}(L/a)$ then decreases with 
$L/a$, 
accounting for the dependence of $a m_\psi^{\rm sat}(L/a)$ 
on $L/a$ 
in Figure \ref{fig:ps_v_versus_pcac}
\footnote{
 The finite size effect observed here differs from 
the power correction found in Ref.~\cite{Fukugita:1992jj}.
}.
 However, this is actually {\it not}
the case in ${\rm SU(2)_C}$ gauge theory
with two adjoint Dirac fermions, 
where $a m_{\rm P}^{\rm sat}(L/a)$ {\it increases with $L/a$}
\cite{DelDebbio:2011kp}. 
 We recall that
the finite size correction 
consists of two terms \cite{Luscher:1985dn}; 
the term induced by $PP$ scattering which increases 
the pseudoscalar meson mass $a m_P$, 
and the one induced by the propagation of $0^{++}$ which decreases $a m_P$.
 Since the two adjoint fermion system
contains a glueball-rich $0^{++}$ lighter than $P$
\cite{DelDebbio:2010hx},
the latter contribution is possibly larger than the former.
 The decrease of $a m_{\rm P}^{\rm sat}(L/a)$ with $L/a$
in our target system suggests that the $PP$ scattering contribution
is the dominant source of the finite size effect.
 The knowledge on the mass of the lightest $0^{++}$ 
will surely give us more comprehensive understanding on this issue.

 Our interest is
if the property (\ref{item:Degeneracy_P_V}) reflects
the dynamics of two-color QCD with 
six flavors of almost massless quarks.
 The high degeneracy in masses between vector and pseudoscalar mesons
was also observed in the ${\rm SU(2)_C}$ gauge theory with two adjoint
fermions \cite{DelDebbio:2011kp}, 
which is considered to be conformal in the deep IR.
 Meanwhile, such a degeneracy 
reminds us the spectrum of bound states 
of massive quarks, $m_q \gg \Lambda_{\rm QCD}$. 
 Another possibility is that
it may occur as a consequence of finite size effect; 
even if the theory is confining, 
the system size is too small for the confining force to 
act between quark and anti-quarks so that they are bounded
solely by $\mathbb{Z}$-copies of Coulombic forces.
 The knowledge on the heavy quark potential will help to answer these questions.

\begin{figure}[htb]
 \begin{minipage}[t]{.48\textwidth}
  \includegraphics[width=7.3cm,clip]{\dirFig/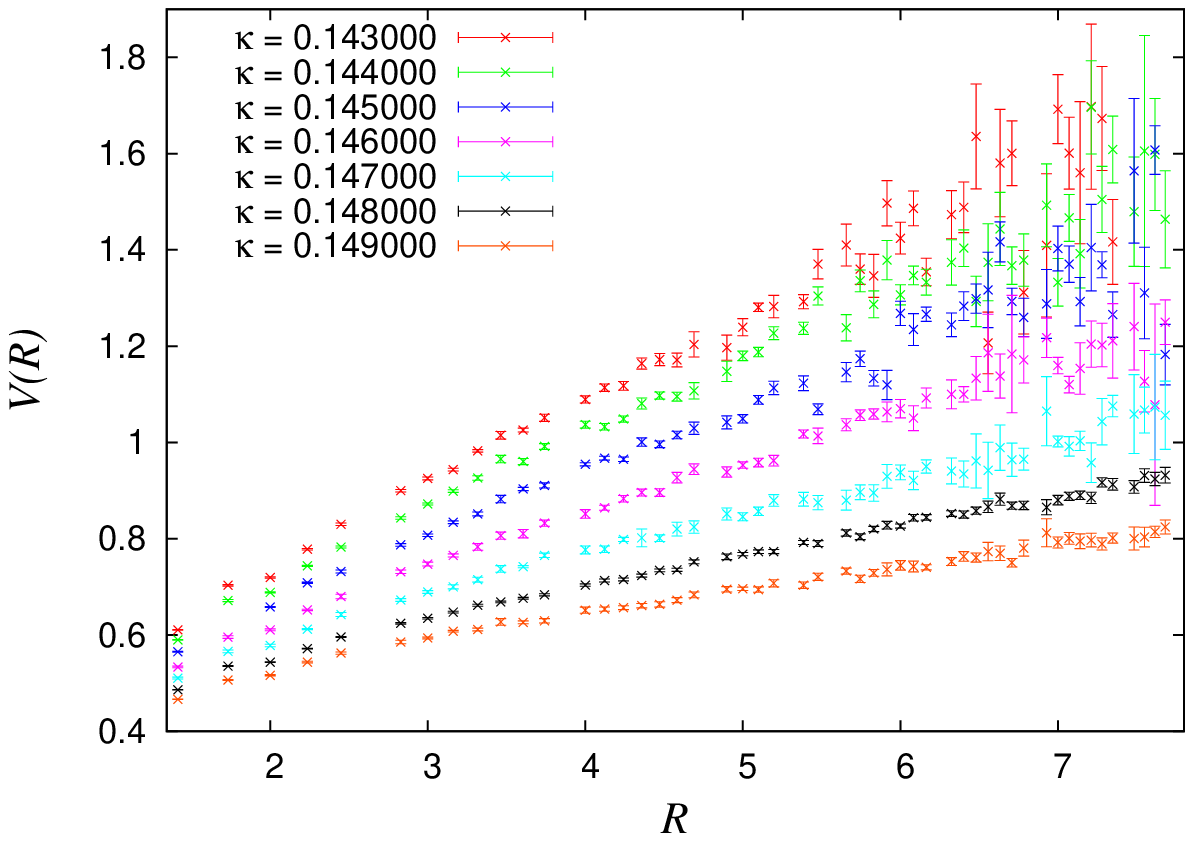}
 \end{minipage}
 \hfill
 \begin{minipage}[t]{.48\textwidth}
  \includegraphics[width=7.3cm,clip]{\dirFig/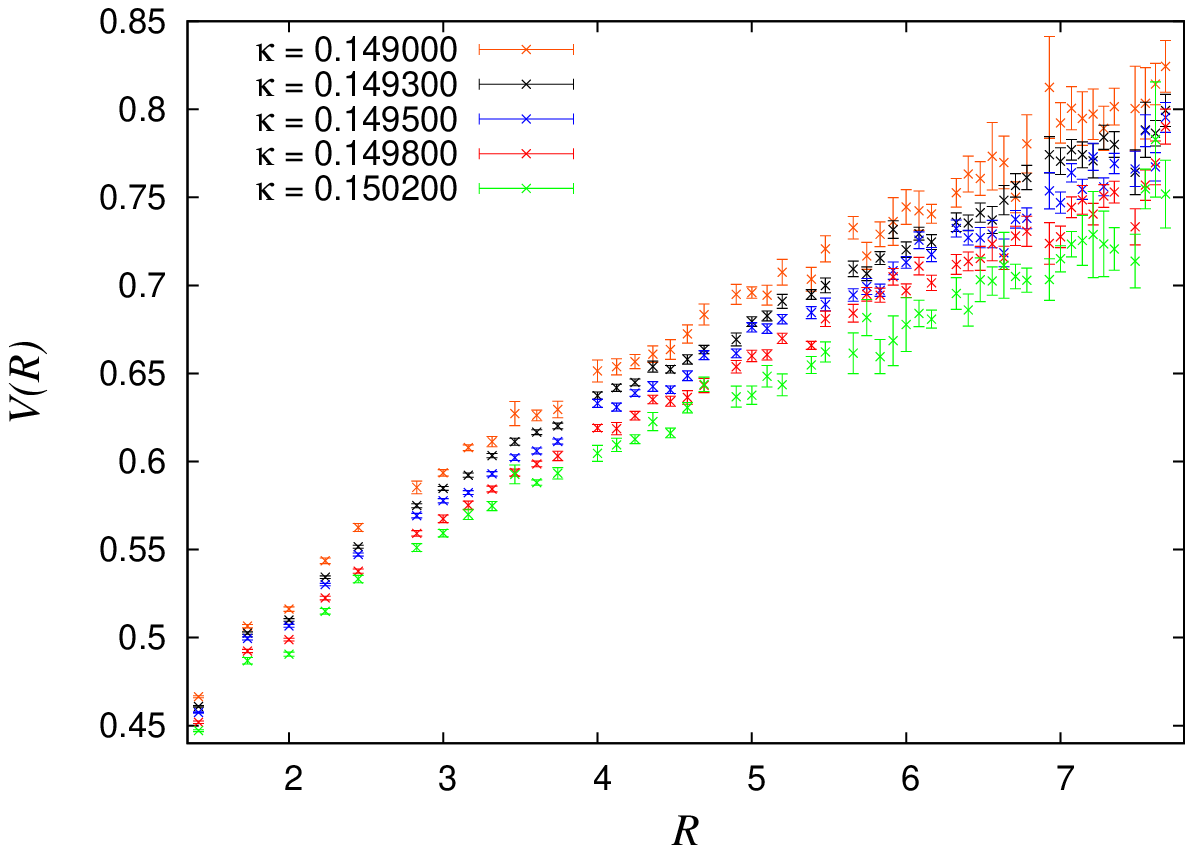}
 \end{minipage}
  \caption{
   Heavy quark potential obtained from
   the lattice with $24^3 \times 48$, $\beta=2.0$,
   $0.1490 \le \kappa \le 0.1498$ (left panel), 
   and $0.1498 \le \kappa \le 0.1502$ (right panel).
   Note that the scale of the ordinate of the right differs from
   that in the left.
 }
 \label{fig:pot_24x24x24x48_b2.0}
\end{figure}

\begin{wrapfigure}[14]{r}{8.0cm}
\begin{center}
\raisebox{-11.0em}[0pt][11.0em]{
  \includegraphics[width=8.0cm,clip]{\dirFig/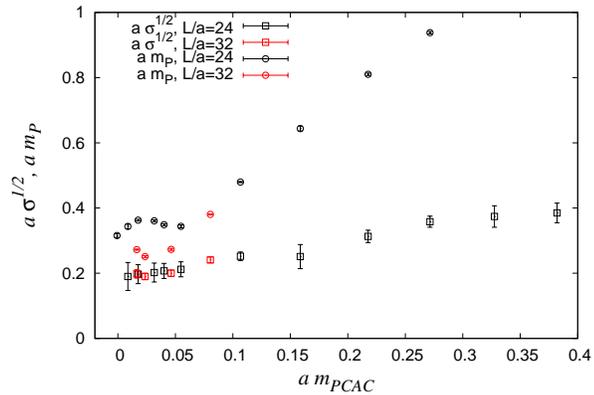}
}
\caption{Masses of lightest pseudoscalar (P) and squared root of string
tension $a\sqrt{\sigma}$
.}
\label{fig:PSandRTensionVesusPCACmass}
\end{center}
\end{wrapfigure}

 The heavy quark potential is extracted from 
the measurement of various sizes of Wilson loops.
 To reduce short distance fluctuation, 
four-level Wilson flow smearing \cite{Luscher:2009eq}
with the weight $\epsilon = 0.01$ in the exponent
was performed for link variables in the spatial directions.
 Figure \ref{fig:pot_24x24x24x48_b2.0} plots 
the heavy quark potentials 
on the lattices with $L/a=24$ and various $\kappa$.
 See Figure \ref{fig:kappaInv_versus_pcac} for the correspondence 
of $1/\kappa$ and $a m_{\rm PCAC}$.
 We can see that the heavy quark potential contains 
the component of linear term, \emph{i.e.},
confinement.

 Figure \ref{fig:PSandRTensionVesusPCACmass} shows
the squared root $a \sqrt{\sigma}$ of the string tension.
 The comparison of $a \sqrt{\sigma}$ for $L/a=24$ 
with that for $L/a=32$ indicates that the finite size effect
is not substantial for string tension. 
 Moreover, there is no such a supporting evidence that
it approaches to zero in the limit $a m_{\rm PCAC} \rightarrow 0$,
suggesting that 
the theory is confining in the chiral limit.
 However, $a m_{\rm P}$ relative to $a \sqrt{\sigma}$
in Figure \ref{fig:PSandRTensionVesusPCACmass} cautions
that if the system exhibits spontaneous chiral symmetry breakdown,
the simulation on the lattice, say with size $L/a=48$, 
will be necessary to capture genuine chiral dynamics.

\begin{figure}[htb]
 \begin{minipage}[t]{.48\textwidth}
  \includegraphics[width=7.5cm,clip]{\dirFig/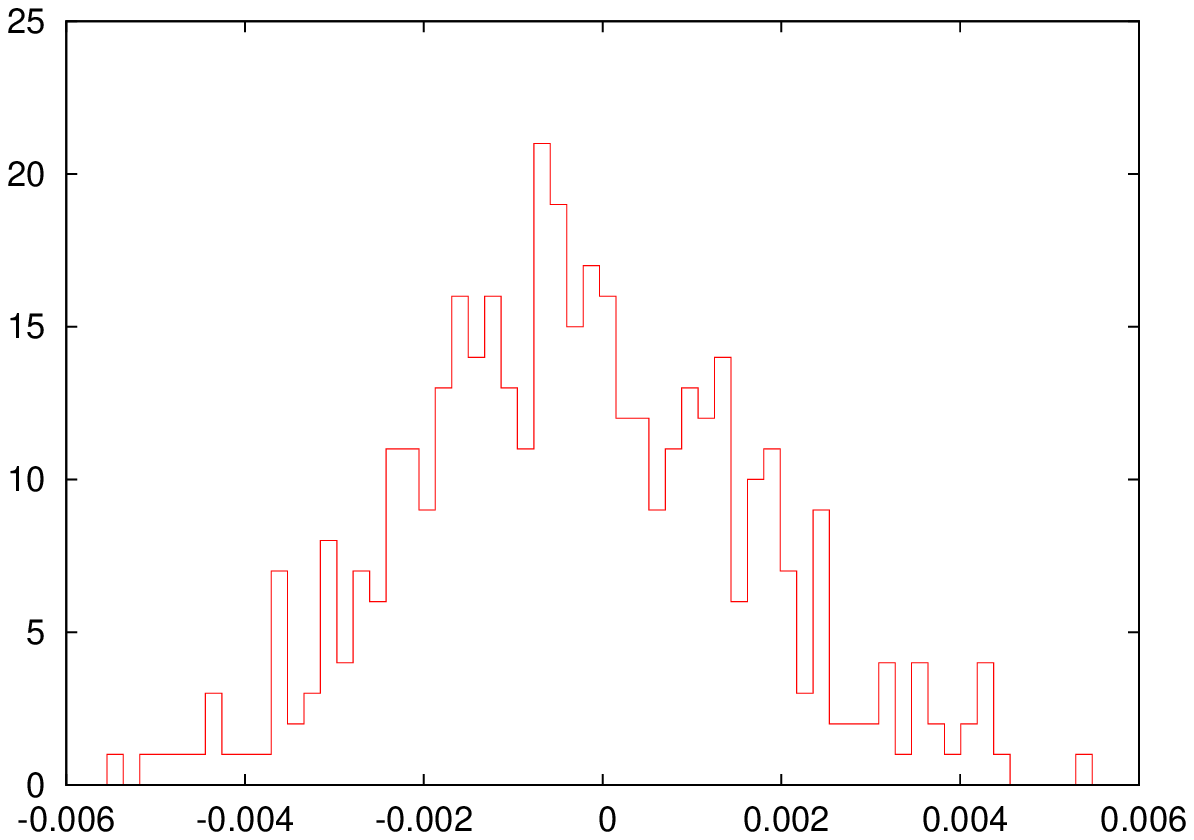}
  \caption{
   Distribution of Polyakov loops in the $z$-direction ($N_Z = 32$, periodic)
   on the lattice with $32^3 \times 64$, $\beta=2.0$, $\kappa=0.14965$.
  }
 \label{fig:ploop_dist_32x32x32x64_b2.0}
 \end{minipage}
 \hfill
 \begin{minipage}[t]{.48\textwidth}
  \includegraphics[width=7.5cm,clip]{\dirFig/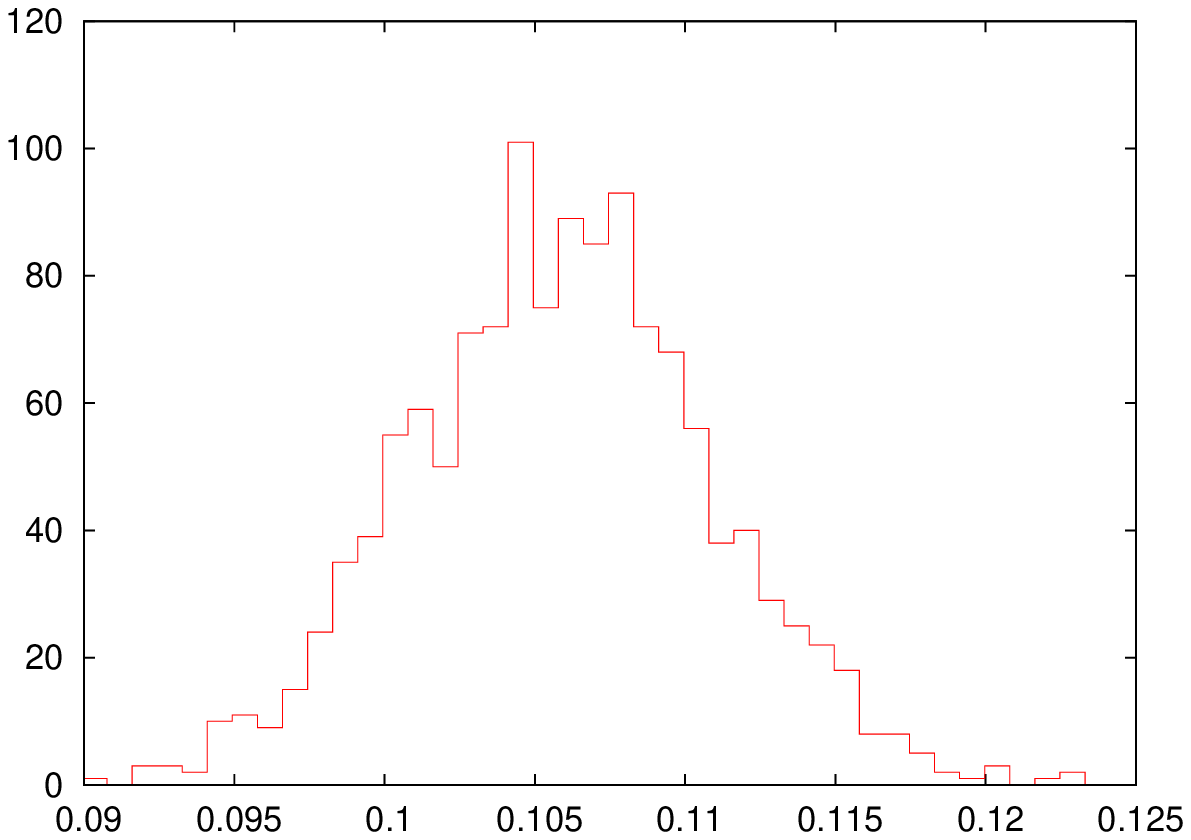}
  \caption{
   Distribution of Polyakov loops in the direction 
   ($N_T = 8$ and anti-periodic)
   on the lattice with $18^2 \times 48 \times 8$, $\beta=2.0$,
   $\kappa=0.1492$.
  }
  \label{fig:ploop_dist_18x18x48x8_b2.0_k0.1492}
 \end{minipage}
\end{figure}

 Lastly, we study the distribution of Polyakov loop $P$ 
in the thermal direction. 
 $P$ takes its value in a real number in the ${\rm SU(2)}$ gauge theory. 
 Its vacuum expectation value $\left<P\right>$ 
is not a good measure of confinement/deconfinement transition
in presence of dynamical quarks in the fundamental representation, 
which explicitly breaks the center $\mathbb{Z}_2$.
 In spite of this fact, its distribution will help to capture
the essence of the dynamics;
the system is deconfining if the distribution of $P$ does not cover $0$ 
and is completely asymmetric with respect to the origin, 
while it is confining if it peaks at the origin and 
distributes almost symmetrically.

 We can see that the periodic boundary condition is equivalent 
to the anti-periodic boundary condition
in ${\rm SU(2)_C}$ gauge theory if all matter fields belong to
the representations of $\mathbb{Z}_2$-odd conjugacy classes.
(Only the sign of a $\mathbb{Z}_2$-odd observable depends on
the condition.)
 Thus, we regard one of the spatial directions with
$32$ sites on lattices of $32^3 \times 64$ 
as a thermal direction and measure $P$ along this direction.
 Figure \ref{fig:ploop_dist_32x32x32x64_b2.0} shows 
the distribution of $P$ for $\kappa=0.14965$.
 The periodic/anti-periodic boundary condition 
will reduce the dynamical modes
so that the finite size effect will act to order the system
($\left<P\right> \ne 0$, \emph{i.e.}, 
tendency of deconfinement).
 Even though the aspect ratio is $32/32$, 
the distribution of $P$ in Figure \ref{fig:ploop_dist_32x32x32x64_b2.0}
exhibits disordering, suggesting confinement.
 In contrast, the distribution of $P$ for $N_T = 8$
in Figure \ref{fig:ploop_dist_18x18x48x8_b2.0_k0.1492}
does not cover $0$, suggesting deconfinement.
 The question whether this transition is really thermal type
needs further study.

 To summarize, we show some evidence supporting
confinement in the two-color QCD with six-flavors of quarks.
 As can be seen in the inset of Figure \ref{fig:ps_v_versus_pcac},
the mass splitting between pseudoscalar meson and vector meson
increases gradually in the small quark mass region
until saturation is encountered. 
 The study with larger size of lattices is necessary 
to approach smaller $m_{\rm P} / \sqrt{\sigma}$ and to
get a definite conclusion.
 Now, it is interesting to start the calculation
of the other physical quantities, 
such as the mass of the lightest particle
in the $0^{++}$ channel relative to the decay constant.

\section*{Acknowledgments}
 We thank A.~Patella and L.~Del Debbio for discussion on their work 
in \cite{DelDebbio:2011kp}.
 This work is supported in part by
Grants-in-Aid for Scientific Research
(S)22224003, (C)20540261, (A)22244018, (C)24540276, 
Grant-in-Aid for Young Scientists (B)23740177, (B)22740183, 
and Grant-in-Aid for Scientific Research on Innovative Areas 
20105001, 20105002, 20105005, 23105707.
 The computation on the large lattices ($L/a=16,\,18,\,24$)
was conducted on the supercomputer system $\varphi$
at Nagoya University, and on the servers with GPGPUs at KEK.


\end{document}